\documentclass[twocolumn,showpacs,preprintnumbers,amsmath,amssymb,prl]{revtex4-1}
\usepackage{amsfonts}
\usepackage[english]{babel}
\usepackage[T1]{fontenc}
\usepackage{times}
\usepackage{mathrsfs}
\usepackage{graphicx}% Include figure files
\usepackage{dcolumn}% Align table columns on decimal point
\usepackage{bm}% bold math
\usepackage[colorlinks,bookmarks=true,citecolor=blue,linkcolor=red,urlcolor=blue]{hyperref}
\usepackage[tight, FIGTOPCAP, hang, raggedright, nooneline]{subfigure}

\newcommand\ket [1] {|#1 \rangle }

\newcommand{\av}[1]{\langle #1\rangle}
\newcommand{\bb}[1]{\mathbf{#1}}

\usepackage{color}
\begin{document}

\title{Topology and Interactions in a Frustrated Slab: Tuning from
Weyl Semimetals \\  to ${\mathcal C}>1$ Fractional Chern Insulators}

\author{E.J. Bergholtz$^1$}
\author{Zhao Liu$^2$}
\author{M. Trescher$^1$}
\author{R. Moessner$^3$}
\author{M. Udagawa$^4$}
\affiliation{$^1$Dahlem Center for Complex Quantum Systems and Institut f\"ur Theoretische Physik, Freie Universit\"at Berlin, Arnimallee 14, 14195 Berlin, Germany}
\affiliation{$^2$Department of Electrical Engineering, Princeton University, Princeton, New Jersey 08544, USA}
\affiliation{$^3$Max-Planck-Institut f\"ur Physik komplexer Systeme, N\"othnitzer Stra\ss e 38, D-01187 Dresden, Germany}
\address{$^4$Department of Applied Physics, University of Tokyo, Hongo 7-3-1, Bunkyo-ku, Tokyo 113-8656, Japan}
\
\date{\today}

\begin{abstract}
We show that, quite generically, a [111] slab of spin-orbit coupled pyrochlore lattice exhibits surface states
whose constant energy curves take the shape of Fermi arcs, localized to different surfaces depending
on their quasi-momentum. Remarkably, these persist independently of the existence of Weyl points in the
bulk. Considering interacting electrons in slabs of finite thickness, we find a plethora of known fractional Chern insulating
phases, to which we add the discovery of a new higher Chern number state which is likely a generalization of the
Moore-Read fermionic fractional quantum Hall state. By contrast, in the three-dimensional limit, we argue for the absence of gapped states of the flat surface band due to a topologically protected coupling of the surface to gapless states in the bulk. We comment on generalizations as well as experimental perspectives in thin slabs of pyrochlore iridates.
\end{abstract}

\maketitle

{\it Introduction.---}
The prediction \cite{topo1,topo2,topo3,topo4,topo5} and subsequent experimental observation  \cite{topoexp1,topoexp2} of topological insulators has fundamentally revolutionized the understanding of electronic states of matter during the past decade \cite{toprev1,toprev2,andreibook}. New frontiers in this field include gapless topological phases such as three-dimensional Weyl semimetals \cite{volovikbook, murakami,weylpyro, burkov,Weylreview} exhibiting exotic Fermi arc surface states \cite{weylpyro,Weylreview,teemu,Yamaji}, interaction effects on the gapless surface of topological insulators \cite{flatsurface,surfaceFQH,surfaceFQH2,surfaceFQH3,surfaceFQH4}, and strongly correlated phases akin to fractional quantum Hall states in two-dimensional (2D) lattices (see Refs.~\cite{Emilreview,otherreview} and  references therein). Drawing additional inspiration from the rapid development of growth techniques in fabricating high quality slabs/films/interfaces of oxide materials \cite{oxide}, this work provides intriguing connections between these seemingly disparate frontiers.

\begin{figure}[h!]
\includegraphics[width=0.99\linewidth]{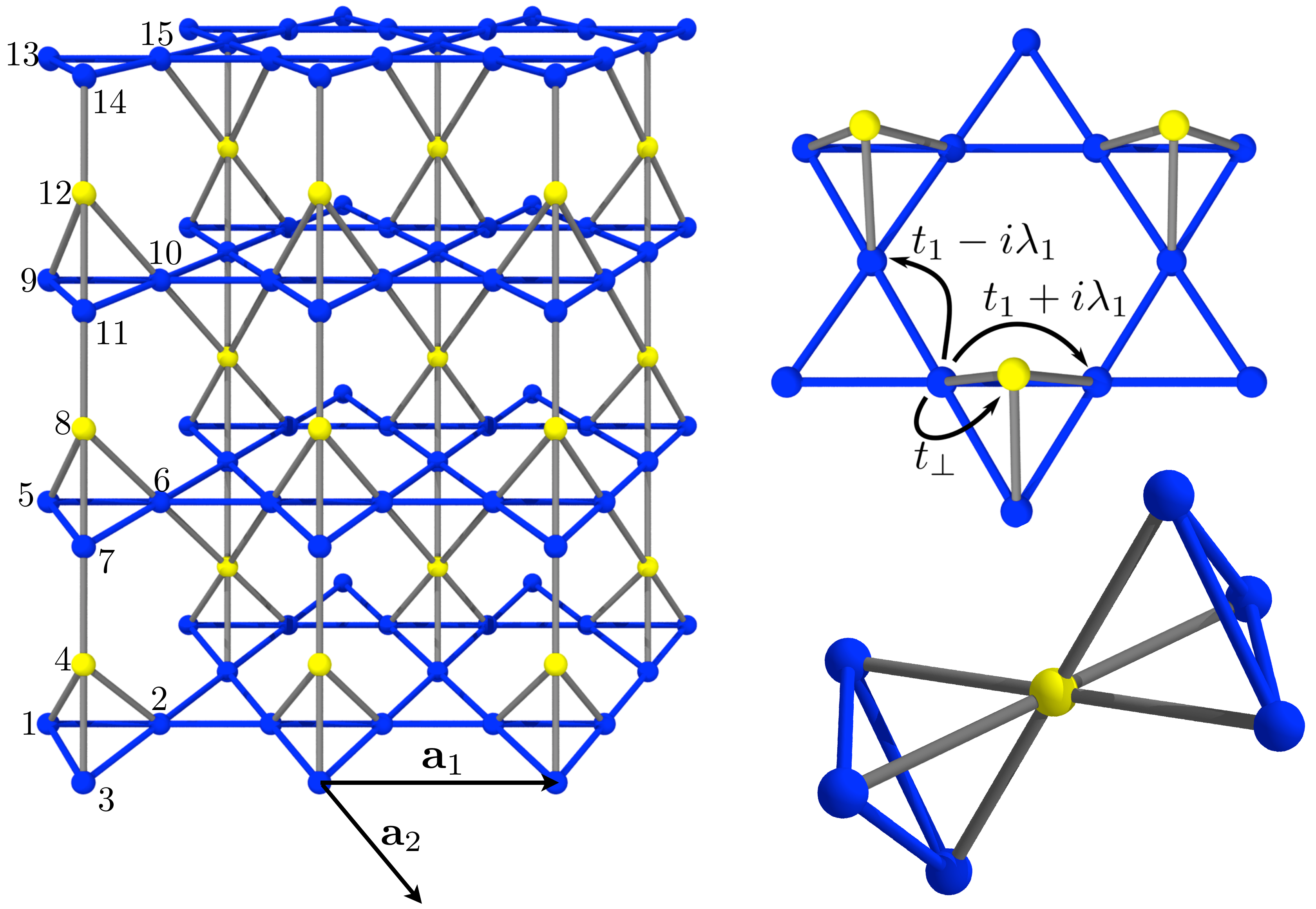}
\caption{{\bf The pyrochlore slab}. The left panel shows the [111] pyrochlore slab  with $N=4$ kagome layers (blue) separated by (yellow) sites of $N-1=3$ triangular layers. A practical labeling of the $4N-1=15$ sites in the unit cell and the basis vectors, $\bb a_1, \bb a_2$, of the Bravais lattice are also indicated. The top right panel indicates the considered nearest neighbor processes. The lower right panel shows the local environment of a triangular (yellow) site for which the local constraint of destructive interference directly leads to the surface states (\ref{psiNgeneral}) at the heart of this work.}
\label{fig:lattice}
\end{figure}

The materials pursuit for Weyl
semimetals and its relatives is rapidly broadening \cite{DWexp,DWexp2,DWexp3,DWexp4}, with spin-orbit
coupled pyrochlore iridates, such as Y$_2$Ir$_2$O$_7$ \cite{weylpyro,max, fiete,nagaosa}
being particularly promising compounds---as these are
favourably grown/cleaved in the [111] direction,  and given their predicted rich variety of strongly
correlated phases \cite{ChernN,ChernN2}, we here study the surface bands of pyrochlore [111] slabs, where
the system can be seen as a layered structure of alternating kagome and triangular layers \cite{max} (Fig. \ref{fig:lattice}).

%%%%%%%%%%%%%%
%%%%%%%%

Our work uncovers an intriguing dichotomy between bulk and surface states which allows us to establish
connections between apparently disparate topological phenomena. While the
bulk band structure changes drastically as a function of the inter-layer tunneling strength $t_\perp$---including
the (dis)appearance of the Weyl semimetal---the surface states, which involve only
the kagome layers, {\it remain unchanged} on account
of their essentially geometrical origin. Most saliently, in the two distinct regimes of $N$  weakly coupled kagome layers, each with unit Chern number, at small $t_\perp$,
and the genuinely three-dimensional Weyl semimetal at large $t_\perp$, identical surface states carrying
Chern number ${\cal C}=N$ are localized at opposite surfaces depending on their momentum.
Constant energy contours in reciprocal space are Fermi arcs, which thus exist also in absence of Weyl
nodes in the bulk!

Upon adding interactions to a partially filled surface band---even when these are made very flat by tuning hopping
parameters---we argue that interactions do not open a gap for thick slabs,
due to a leakage into the bulk along "soft" lines related to projections of remnant Weyl nodes.
However, in thin slabs we find a plethora of possible fractionalized phases, some of which were discovered
earlier \cite{ChernN,ChernN2} with the implicit assumption of sub-critical inter-layer tunneling. Most prominently,
we provide evidence for a first non-Abelian fermionic fractional Chern insulator (FCI) in a ${\cal C}>1$ band, namely a ${\cal C}=2$ generalization
of the Moore-Read quantum Hall phase \cite{mr}. Our work thus gives a unifying and
fresh perspective on the
intriguing combination of fractionalization and topological surface localization impossible in
strictly two-dimensional systems.

{\it Setup.---}
Our tight binding model on  $N$ kagome layers, $\mathcal K_m$,
alternating with $N-1$ triangular  layers, $\mathcal T_m$ \cite{max} (Fig. \ref{fig:lattice}),
considers spinless, spin-orbit coupled,
fermions with interlayer hopping amplitude $t_\perp$ and
kagome layer (next) nearest hopping amplitudes
$t_1\pm i\lambda_1$ ($t_2\pm i\lambda_2$), where the $-$($+$) sign applies for (anti-)clockwise hopping w.r.t.\
the hexagon on which it takes place. Time-reversal symmetry is absent,
e.g.\ due to an orbital field or spontaneous ferromagnetism.
 \begin{figure}[t!]
\centerline{
\includegraphics[width=0.99\linewidth]{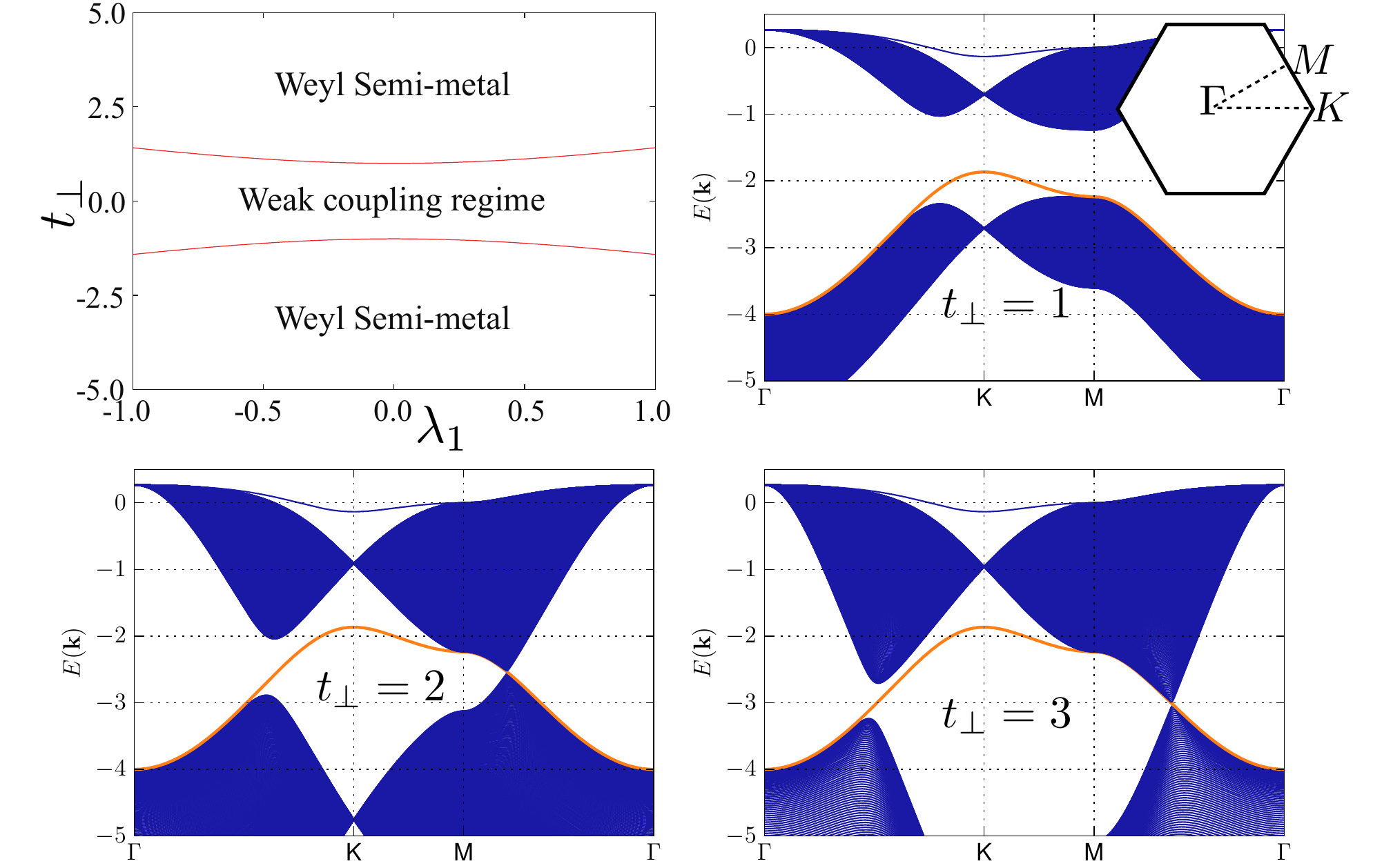}
}
\caption{{\bf Weakly coupled Chern insulators vs. Weyl semimetals}. As $t_\perp$ is increased there is a transition from a weakly-coupled regime to a distinct phase where Weyl nodes occur on the line connecting $\Gamma$ and $M$. In the top left panel we show the phase diagram in the case of nearest neighbor hopping only (we set $t_1=-1$ throughout) \cite{supmat}. The other panels show example band structures with fixed $ \lambda_1=0.5$ and varying $t_\perp=1,2,3$ for a slab with $N=300$ kagome layers along the crucial $\Gamma-K-M-\Gamma$ path  through the projected 2D Brillouin zone (BZ) (cf. top right inset). Note that, remarkably, the band highlighted in orange corresponding to the surface states (\ref{psiN}), is independent of $t_\perp$.
\label{fig:tperp}}
\end{figure}

{\it Band structure and surface wave functions.---}
Independently of the form of the Bloch states of a single kagome layer, three bands of the $N$-layer
system are exactly described by
\begin{eqnarray}
\ket{\psi^i(\mathbf k)}=\mathcal{N}(\bb k) \sum_{m=1}^{N}\Bigr{(}r(\bb k)\Bigr{)}^m\ket{\phi^{i}(\mathbf k)}_m\ ,\label{psiNgeneral}%
\end{eqnarray}
where $\ket{\phi^{i}(\mathbf k)}_m$, $i=1,2,3$ are the single layer Bloch states localized to $\mathcal K_m$ and $\mathcal{N}(\bb k)$ ensures proper normalization. The coefficients $r(\bb k)$ are determined by demanding that the amplitudes for hopping to the triangular layers vanish by interfering destructively (Fig.~\ref{fig:lattice}): $r(\bb k)= -\frac{\phi_1^i(\bb{k}) + \phi_2^i(\bb{k}) + \phi_3^i(\bb{k})}{e^{-i k_2}\phi_1^i(\bb{k}) + e^{i (k_1-k_2)}\phi_2^i(\bb{k}) + \phi_3^i(\bb{k})}$, where $\phi_n^i(\bb{k})$, $n\leq3$, are the components of the
Bloch spinor for the pertinent state $\ket{\phi^{i}(\mathbf k)}$ in a single kagome-layer, and $k_{1,2}=\bb k\cdot \bb a_{1,2}$. While $\phi_n^i(\bb{k})$, $n\leq3$,
can be analytically obtained by diagonalizing $3\times 3$ Hermitian matrices, the full Bloch spinor is
fully known via  $\psi_{4m}^i(\bb{k})=0$, $\psi_{n+4(m-1)}^i(\bb{k})=\mathcal{N}(\bb k)\bigr{(}r(\bb k)\bigr{)}^m \phi_{n}^i(\bb{k})$ for all
$\bb k, n, m$, with $E(\bb{k})$ of the states (\ref{psiNgeneral})  equal to those of the single layer case.

Let us emphasize that, firstly, the states on the slab are exponentially localized to either the top or bottom layers, except in high symmetry cases where $|r(\bb k)|=1$. And secondly, if periodic boundary conditions are applied also in the [111]-direction, there are {\it no} generic eigenstates of the form (\ref{psiNgeneral}), underscoring their surface nature.

\begin{figure}[t!]
\centerline{
\includegraphics[width=0.99\linewidth]{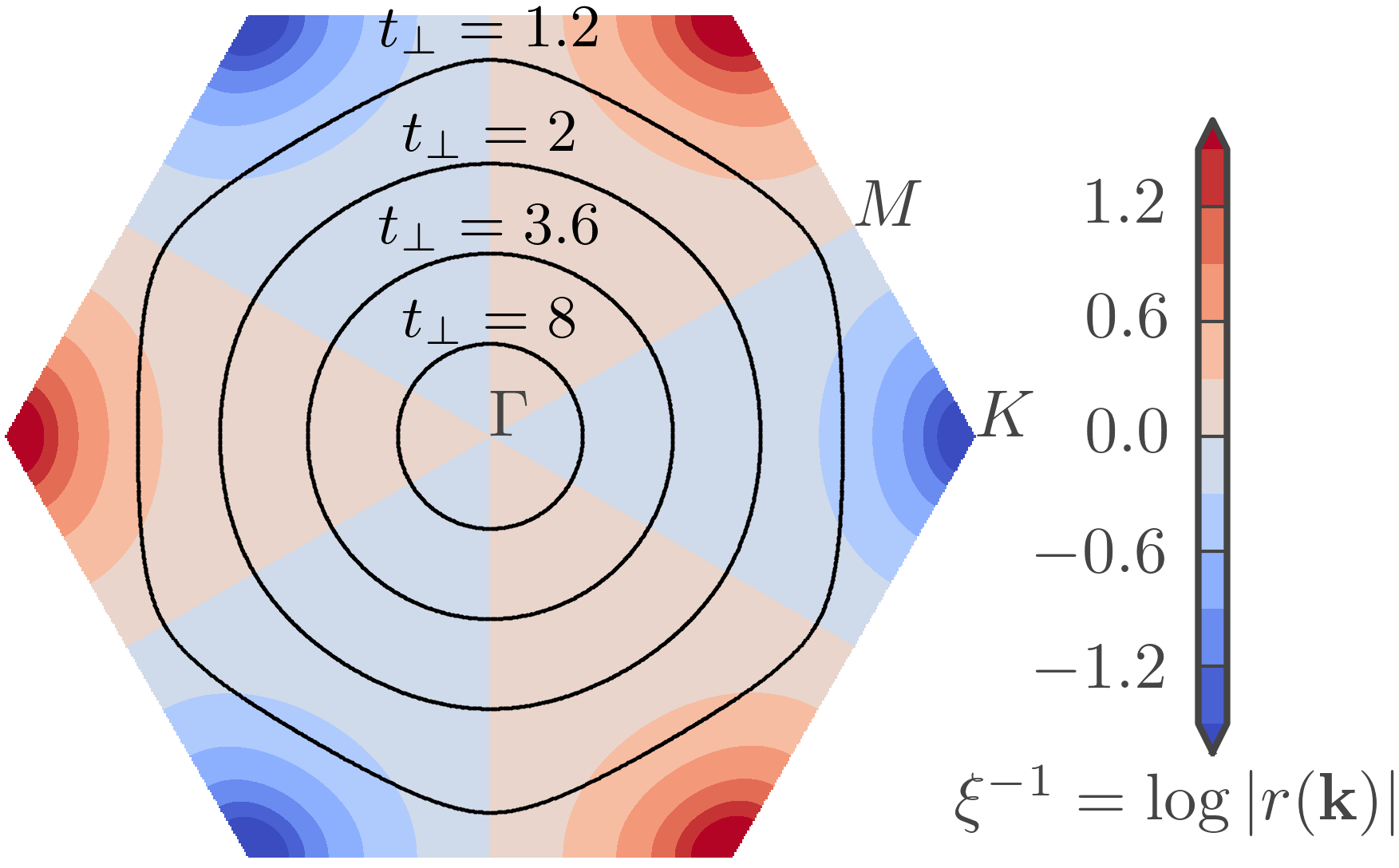}
}
\caption{{\bf Surface state structure and Fermi arcs}. The color scale indicates
the inverse penetration depth, $\xi^{-1}(\mathbf k)=\log(|r(\mathbf k)|)$ of the surface states throughout the
2D BZ for the same parameters, $t_1=-1,\lambda_1=0.5$, used in Fig. \ref{fig:tperp}. The black lines illustrate Fermi arcs for a chemical potential set at the Weyl node for a few $t_\perp$-values. When $\xi^{-1}(\mathbf k)$ changes sign, the localization changes between top (red) to bottom (blue) surfaces, hence splitting the Fermi "circle" into six spatially disjoint arcs.  \label{fig:logrk}}
\end{figure}

\begin{figure}[bt]
\centerline{
\includegraphics[width=0.99\linewidth]{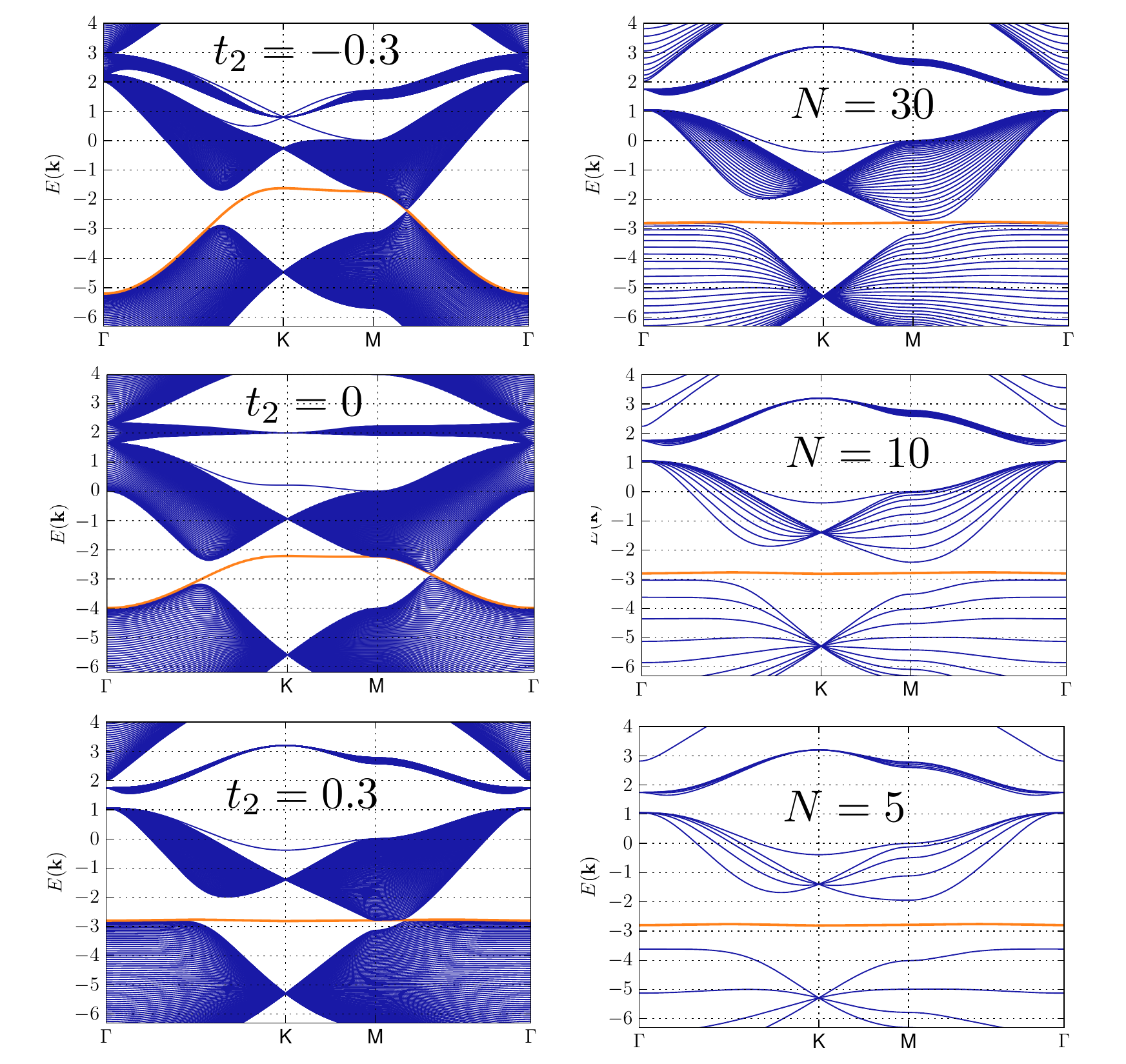}
}
\caption{{\bf From Weyl nodes to flat surface bands}. For $t_1=-1,\lambda_1=0.3,\lambda_2= 0.2,t_{\perp}=2.0$, we plot the energy dispersion for various $N$
and $t_2$ on the path $\Gamma-K-M-\Gamma$ through the BZ.
In the left panel we set the number of kagome layers to $N=100$ while varying $t_2$. For $t_2=-0.3$ (top) there is a clearly visible Weyl node on the line connecting $M$ and $\Gamma$. For $t_2=0$ the node is skewed and at $t_2=0.3$ it is essentially flattened while the surface band (bold orange) remains at almost fixed energy throughout the entire BZ.
In the right panel we fix $t_2=0.3$ and reduce the number of layers, $N=30, 10, 5$, from top to bottom,  and
a sizable finite size gap quickly opens throughout the entire BZ. The band highlighted in bold orange is that of (\ref{psiN}), carrying Chern number ${\cal C}=N$.
}
\label{highCdispersion}
\end{figure}

In the following, we consider the case of single layer kagome bands carrying non-zero Chern number \cite{haldanemodel,chernins1,refref}, say ${\cal C}=1$. Then,
the multilayer state (\ref{psiNgeneral}) has Chern number $N$:
\begin{eqnarray}
\ket{\psi^{{\cal C}=N}(\mathbf k)}=\mathcal{N}(\bb k) \sum_{m=1}^{N}\Bigr{(}r(\bb k)\Bigr{)}^m\ket{\phi^{{\cal C}=1}(\mathbf k)}_m\label{psiN}%
\end{eqnarray}
where $\ket{\phi^{{\cal C}=1}(\mathbf k)}_m$ is the state localized to $\mathcal K_m$. The states (\ref{psiN}) play a prominent role in this work, and their corresponding energies are highlighted in bold orange throughout this work (not shown are the two related states with ${\cal C}=0,-N$).

Fig.~\ref{fig:tperp}  illustrates the finite $t_\perp$ transition between weakly coupled Chern insulators and the Weyl semimetal regime with linear band
touching points described by
\begin{eqnarray}
H_{{\rm Weyl}}= \sum_iv_i \sigma_i k_i+ E_0(\bb k)\mathbb{I}\ , \label{Hweyl}\end{eqnarray}
where $\sigma_i$ are Pauli matrices and $\mathbb{I}$ is the identity matrix. Precisely at the transition, the valence and conduction bands exhibit a two-fold degenerate touching at the $M$-points, which split into three pairs of (non-degenerate) Weyl cones that travel towards the $\Gamma$-point where they meet as $t_\perp\rightarrow\infty$. Remarkably, the states (\ref{psiN}) are entirely independent of the value of $t_\perp$; in each case they describe states localized to the surfaces perpendicular to the [111] cleavage/cut/growth direction, while at the same time their interpretation fundamentally changes.
Note also that the dispersion of the states (\ref{psiN}) always traverses the Weyl point.

At fixed chemical potential, which may be fixed at the Weyl node due to stoichiometric considerations, the states (\ref{psiN}) precisely describe Fermi arcs. In Fig. \ref{fig:logrk} we illustrate the momentum dependence of the surface localization of the states (\ref{psiN}). Most saliently, we find that the penetration depth diverges along the lines connecting $\Gamma$ and $M$. Crossing these lines, the localization changes between the bottom and top surfaces, which is the hallmark behavior of Fermi arcs. More specifically, a typical Fermi "circle" splits
into six Fermi arcs which switch between top and bottom surface six times, whenever
 the Fermi circle crosses a $\Gamma$-$M$-line (cf. Fig. \ref{fig:logrk}).

In Fig. ~\ref{highCdispersion} we go on to show how the state (\ref{psiN}) can smoothly be
transformed into a band which is essentially dispersionless, yet being
tightly attached to bulk bands (for thick slabs). It is important to note that also the latter regime is described by a Weyl Hamiltonian (\ref{Hweyl}) with a suitable choice of $E_0(\bb k)$; the essential point is that the topology is unchanged as long as the band touching is linear ($v_i\neq 0, i=1,2,3$), no matter how skewed the Weyl point is due to the overall constant dispersion $E_0(\bb k)$. In fact, the Weyl nodes carry a quantized Chern flux and can as such only be annihilated by merging with an opposite chirality partner \cite{Weylreview}. Furthermore, considering a quasi-two-dimensional slab, one finds that there is a fairly sizeable region in which the bandwidth is (much) smaller than the band gap, although to obtain very flat surface bands we need to include also next-nearest neighbor hopping (as done in
Fig.~\ref{highCdispersion}). Crucially, this holds true for thin slabs both in the weakly coupled regime studied earlier \cite{max,ChernN,ChernN2}, as well as when the bulk is in the Weyl semimetal regime.

Thus, one can consider the flat bands of Refs.~\cite{max,ChernN,ChernN2} vestiges of
Weyl semimetal surface bands. While $t_\perp$ considered in those works is slightly below the Weyl semimetal regime, our exact solution (\ref{psiN}) reveals that this distinction is in fact immaterial in thin slabs as long as only the topological band is concerned.

{\it Projected interactions in the flatbandlimit.--}
We now add  interactions to a partially filled surface band with ${\cal C}=N$; for Weyl semimetals with the chemical potential pinned to the Weyl node in the bulk by stoichiometry, this may well be relevant to the low-energy physics of quasi-2D slabs.

The matrix elements of {\em any} local interaction (provided it is uniform throughout the lattice and does not couple different kagome layers) follows from (\ref{psiN}); for a two-body interaction,
\begin{eqnarray} V_{\mathbf k_1 \mathbf k_2 \mathbf k_3 \mathbf k_4}^{{\cal C}=N}\! &=&
\! V_{\mathbf k_1 \mathbf k_2 \mathbf k_3 \mathbf k_4}^{{\cal C}=1}\! \Bigr(  \frac{|r(\bb{k}_1)|^2-1}{|r(\bb{k}_1)|^{2N}\!\!  - 1}\! \cdots\!  \frac{|r(\bb{k}_4)|^2-1}{|r(\bb{k}_4)|^{2N}\!\!  - 1} \Bigr)^{\!\frac 1 2}\nonumber\\ &\times&
 \frac{(r^*(\bb{k}_1)r^*(\bb{k}_2)r(\bb{k}_3)r(\bb{k}_4))^N-1}{r^*(\bb{k}_1)r^*(\bb{k}_2)r(\bb{k}_3)r(\bb{k}_4) - 1} \label{layersum}
\end{eqnarray}
where the band projected interaction Hamiltonian in general can be written as
\begin{eqnarray}H_{{\rm int}}=\sum_{{\mathbf k}_1
 \mathbf k_2 \mathbf k_3 \mathbf k_4}\!\!\!
 V_{\mathbf k_1 \mathbf k_2 \mathbf k_3 \mathbf k_4}^{{\cal C}=N}c^{\dagger}_{\mathbf k_1}c^{\dagger}_{\mathbf k_2}c_{\mathbf k_3}c_{\mathbf k_4} \ ,
 \label{hamf}\end{eqnarray}
where $c^{\dagger}_{\mathbf k}$ ($c_{\mathbf k}$) creates (annihilates) an electron in the state $\ket{\psi^{{\cal C}=N}(\mathbf k)}$. These expressions generalize straightforwardly to any local $(k+1)-$body interaction. It is important to note that both the magnitude and the the relative phase factors of the matrix elements depend non-trivially on $N$.

{\it Fractional topological phases.--}
The flat bands in our model \cite{max} are known to host series of FCIs in bands with ${\cal C}>1$:
%; In Ref. \onlinecite{ChernN}
stable Abelian FCIs of fermions at band filling $\nu=\frac 1 {2{\cal C}+1}$ and bosons at
$\nu=\frac 1 {{\cal C}+1}$\cite{ChernN}  with non-Abelian phases of bosons at $\nu=\frac k{{\cal C}+1}$ using onsite $(k+1)$-body interactions \cite{ChernN2}. 

\begin{figure}[h]
\centerline{
\includegraphics[width=\linewidth]{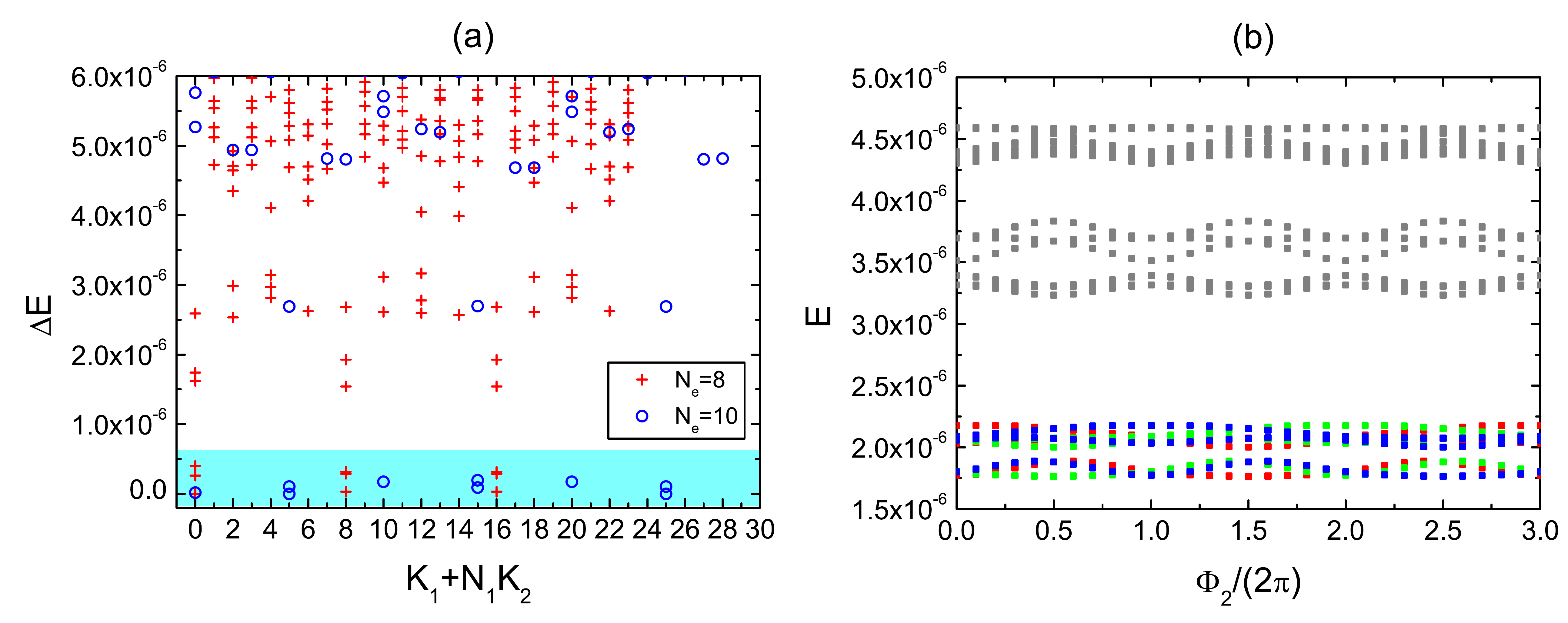}
}
\caption{{\bf Topological degeneracy in the ${\cal C}=2$ bilayer system}. (a) The energy spectra of the interaction $H=\sum_{\langle i,j,k\rangle}n_in_jn_k$ interaction projected to the flatbandfor $N_e=8$ and $N_e=10$ electrons in lattices with $N_{1}\times N_{2}=(N_e/2)\times 6$ unit cells yielding a filling fraction of $\nu=1/3$. Each energy level is labeled by the conserved many-body momentum $(K_1,K_2)$. The shaded area indicates the $9$ quasi-degenerate states. (b) The $y$-direction spectral flow for $8$ electrons under twisted boundary conditions $\Psi(\mathbf{r}_j+N_{2}\mathbf a_{2})=\exp(i\Phi_2)\Psi(\mathbf{r}_j)$ of the ground state $\Psi(\mathbf{r}_j)$. The red, green and blue dots represent the $9$ quasi-degenerate states in different momentum sectors, and the gray dots represent the excited states. The parameters are $t_1=-1,\lambda_1=0.9,t_2=\lambda_2=0$. \label{fig:MR}}
\end{figure}

Here, we identify a new FCI at filling fraction $\nu=1/3$, which we
propose as a candidate for the first non-Abelian Fermionic FCI for ${\cal
C}=2$, a generalization of the Moore-Read quantum Hall state \cite{mr}. This is based on its large and robust 9-fold
topological degeneracy (Fig. \ref{fig:MR}), its
non-trivial entanglement spectra  \cite{supmat} as well as its provenance from
a three-body interaction (see \cite{supmat} for details). We note that, given the
flat  band is not located at the bottom of the spectrum, this state
is indeed naturally suited to fermions at an appropriate density.

{\it Gapless bulk.--}
Next, we argue that generically, a flat surface band will not be
gapped by interactions within it for thick slabs. This
happens because of the -- topologically stable -- locations on lines
in reciprocal space where the states of the band switch the surface at
which they are localised. At these points, the inverse penetration
depth $\xi^{-1}(\mathbf k)= \log(|r(\mathbf k)|)$
(cf. Fig. \ref{fig:logrk}) vanishes. Matrix elements $V_{\mathbf k_1
  \mathbf k_2 \mathbf k_3 \mathbf k_4}^{{\cal C}=N}$ involving
$n_{\{k_i\}}$ momenta on these lines vanish as $\big(\frac 1
{\sqrt{N}}\big)^{n_{\{k_i\}}}$, reflecting the spatial spread of the
wavefunctions.

\begin{figure}[h]
\centerline{
\includegraphics[width=0.99\linewidth]{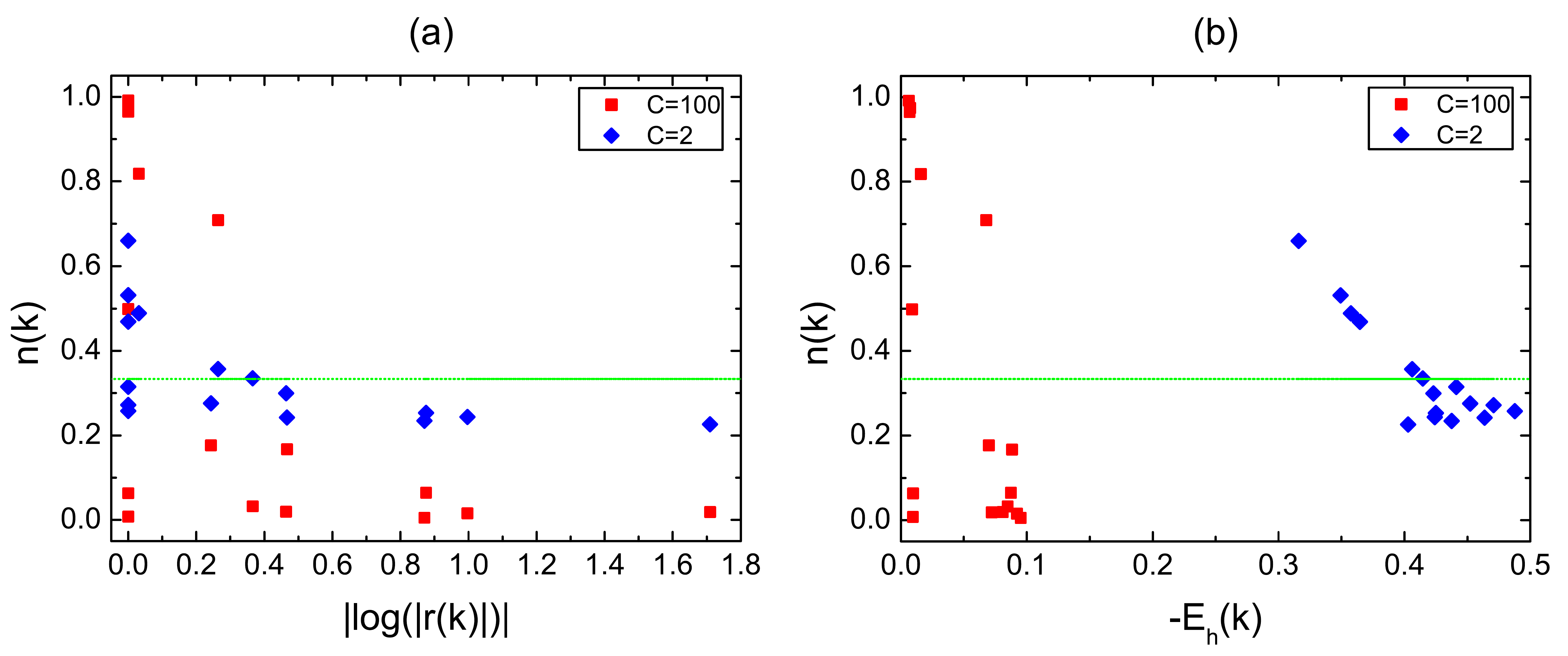}
}
\caption{{\bf Ground state occupation numbers}. $n(\bb k)$ plotted against $|\xi(r(\bb k))|^{-1}=|\log (|r(\bb k)|)|$ (left panel) and $-E_h(\bb k)$ (right panel, cf. Ref. \cite{andreas}) for Chern numbers ${\cal C}=2$ (blue diamond) and ${\cal C}=100$ (red square) at $\nu=1/3$ . This illustrates a general trend: $n(\bb k)$ generically very inhomogenous at large ${\cal C}$ while it can remain comparably constant for small ${\cal C}$ (and $\nu$) \cite{supmat}.  \label{fig:occ}}
\end{figure}

This is borne out by our numerics, where the absence of a FCI is
indicated by an inhomogeneous electron distribution $n(\bb k)$ in
reciprocal space, reminiscent of a Fermi surface
(Fig. \ref{fig:occ}). This is analogous to the compressible states at
high filling fractions in ${\cal C}=1$ bands, where an effective "hole-dispersion", $E_h(\bb k)$  \cite{supmat}, resulting from a particle-hole transformation, dictates the low-energy physics \cite{andreas}. In fact, both $\xi^{-1}(\mathbf k)$
and $E_h(\bb k)$ correlate rather well with $n(\bb k)$ as illustrated
for $\nu=1/3$ with nearest neighbor repulsion,
$H=\sum_{\av{i,j}}n_in_j$ (Fig. \ref{fig:occ}, see the
supplementary materials for details and further examples \cite{supmat}).

{\it Discussion.---} In this work, we have unraveled a striking
connection between seemingly distinct frontiers of contemporary
condensed matter physics by explicitly demonstrating that flat
bands with Chern number ${\cal C}=N$ appearing on a slab of pyrochlore
\cite{max}, and known to harbor a rich variety of fractional Chern
insulators \cite{ChernN,ChernN2}, are in fact surface state vestiges
of the Fermi arcs of Weyl semimetals \cite{weylpyro}. This result has a bearing in the general context of the
bulk-boundary correspondence in topological matter: while it has been realized that there can be phase transitions on the boundary while leaving the bulk intact  \cite{bulkboundary1,bulkboundary2}, we find a striking example of the converse situation with a bulk transition leaving the boundary theory unaffected. 

We note that layered structures, albeit with a rather different
alternating normal insulator-strong topological insulator setup, have
been suggested earlier as a possible platform for Weyl semimetals
\cite{burkov,burkov2}. Weyl semi-metals have also been predicted to
occur in pyrochlore based bulk materials, in particular in
A$_2$Ir$_2$O$_7$ (A is a rare-earth element) iridates \cite
{weylpyro}, for which the existence of remnant Fermi arc states at
certain magnetic domain walls even in the absence of bulk Weyl nodes
was recently suggested \cite{Yamaji}. Given the experimental
advantages with finite pyrochlore slabs grown in the [111] directions \cite{fiete,nagaosa,max}, as compared to
other oxide interfaces such as perovskite heterostructures
(which may also harbor intriguing flat bands \cite{digital,c2}), and the generality of our exact solutions for the
surface states based solely on locality and lattice geometry, our setup has its distinct advantages even before
considering intricate interaction effects.

The exact solutions (\ref{psiNgeneral},\ref{psiN}) provide a generic
recipe for "engineering" exotic surface states: coupling Chern insulators
with a desirable, e.g. flat, dispersion \cite{kapit} in a geometrically frustrated manner results in states with the same dispersion but with higher Chern number and added complexity of Fermi arc variety. While we focused on pyrochlore slabs, this procedure generalizes to other frustrated lattices.

We have also explored the effect of interactions in these bands and identified new
fractionalized topological phases as well as generic gapless states as ${\cal C}\rightarrow\infty$. Our work establishes that the combined fractionalization and topological surface localization of the interacting states found here,
and in Refs.~\cite{ChernN,ChernN2}, are impossible in strictly two-dimensional (isotropic) models just as Fermi arcs cannot exist in
purely two-dimensional band materials. This feature distinguishes the pyrochlore based FCIs from other ${\cal C}>1$ generalizations of multilayer quantum Hall states \cite{disloc,ChernTwo,dassarma,layla,Grushin,yangle,zhao2,sunnew,masa}.

The present work invites a number of interesting questions regarding the interplay between fractionalization, surface localization and translation symmetry. In this context, it would be particularly interesting to investigate the effects of lattice dislocations \cite{disloc,dislocweakTI}.
\\
\acknowledgments
We thank J. Behrmann and A. L\"auchli for related collaborations. EJB
is grateful to P. Brouwer, T. Ojanen and B. Sbierski for numerous
discussions on Weyl semi-metals.  EJB and MT are supported by
DFG\textquoteright{}s Emmy Noether program (BE 5233/1-1). Z.~L.~is
supported by the Department of Energy, Office of Basic Energy
Sciences, through Grant No.~DE-SC0002140. M. U. is supported by Grants-in-Aid for Scientific Research (No. 24340076, 26400339, and 24740221). This work was in part supported
by the Helmholtz VI ``New States of Matter and Their Excitations''.

\setcounter{equation}{0}
\setcounter{figure}{0}
\renewcommand{\theequation}{S\arabic{equation}}
\renewcommand{\thefigure}{S\arabic{figure}}
\renewcommand{\bibnumfmt}[1]{[S#1]}
\renewcommand*{\citenumfont}[1]{S#1}

\section*{Supplementary material}
In this supplementary material we present a more exhaustive single particle phase diagram of the pyrochlore slab model, and provide additional details and data for the interacting flat band problem to illustrate the contrasting behavior in thin and thick slabs. Our data  corroborate our conclusion of the absence of insulating states in the flat surface bands of very thick slabs, i.e. at very large Chern numbers, where the occupation numbers $n(\bb k)$ are generically inhomogeneous and feature Fermi surface-like kinks, which are characteristic of the predicted gapless states. Finally, we present entanglement spectra and some general arguments in favor of the non-Abelian interpretation of the discovered $\nu=1/3$, $\mathcal{C}=2$ fractional Chern insulator.

\subsection*{Single particle phase diagram}

\begin{figure}
\includegraphics[width=\linewidth]{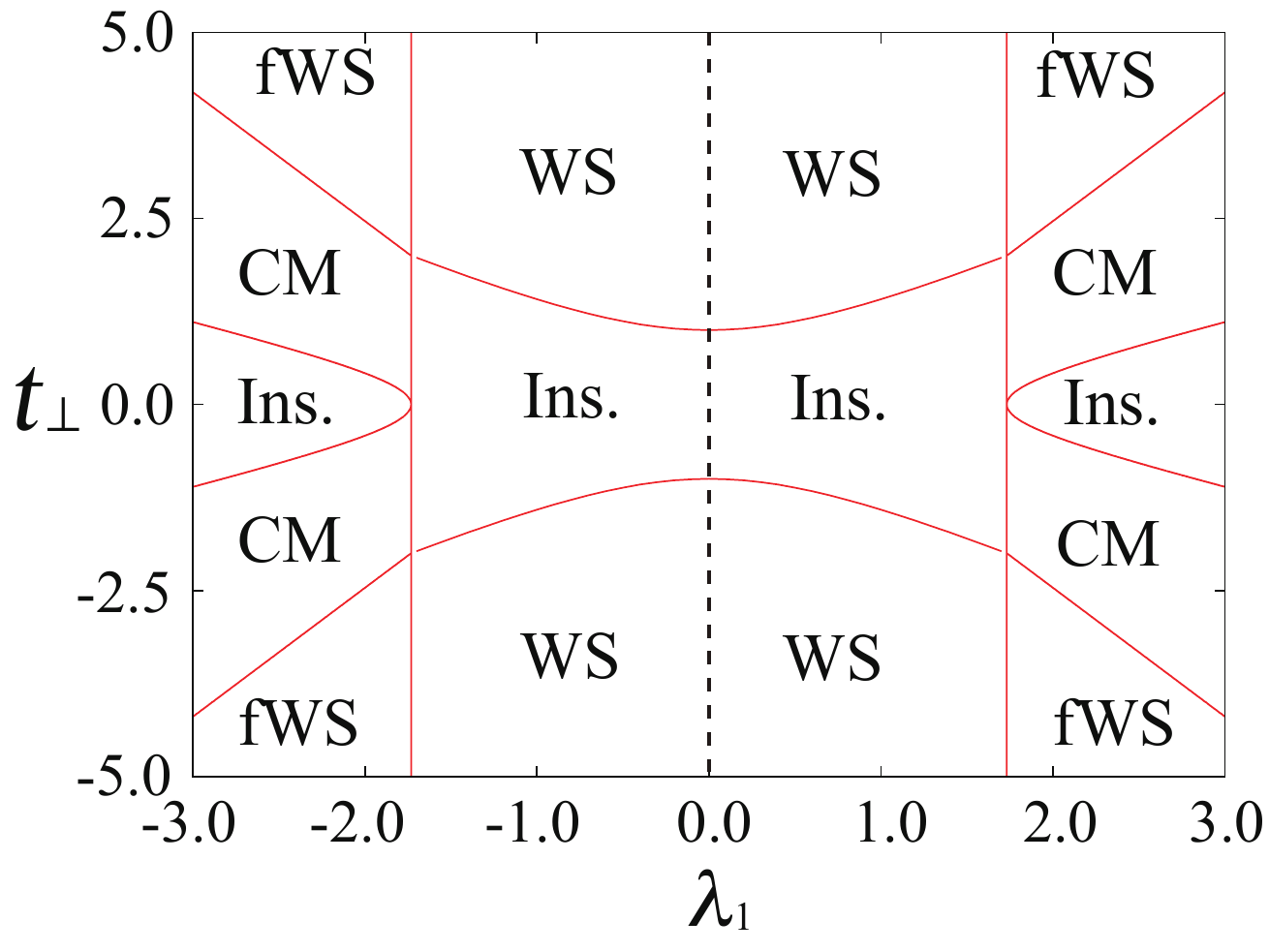}
\caption{The complete nearest-neighbour phase diagram of the pyrochlore slab model \cite{maxS}, which is composed of the insulating region (Ins.), the Weyl semi-metal phase (WS), compensated metallic phase (CM), and the flat Weyl semi-metal phase (fWS). For the characterization of fWS phase, see the text below. The dashed line at $\lambda_1=0$ means the line degeneracy ("line node") remains along this line.}
\label{fig:phased}
\end{figure}

We begin this supplementary material by providing a more complete single particle phase diagram for the pyrochlore slab model \cite{maxS}, as shown in Fig. \ref{fig:phased}. Compared to the phase diagram presented in the main text, this phase diagram covers a wider range of parameter values,  and despite the simplicity of the model, we find that the phase diagram has a quite rich structure. In total there are five different bulk phases present for generic nearest neighbor parameters, $t_\perp, \lambda_1$ (we keep $t_1=-1$ fixed).
Each phase has distinct nature of bulk single-particle excitation gaps at quarter filling. The excitation gap is obtained by assuming periodic boundary conditions for all three directions.

\begin{figure*}
\includegraphics[width=\linewidth]{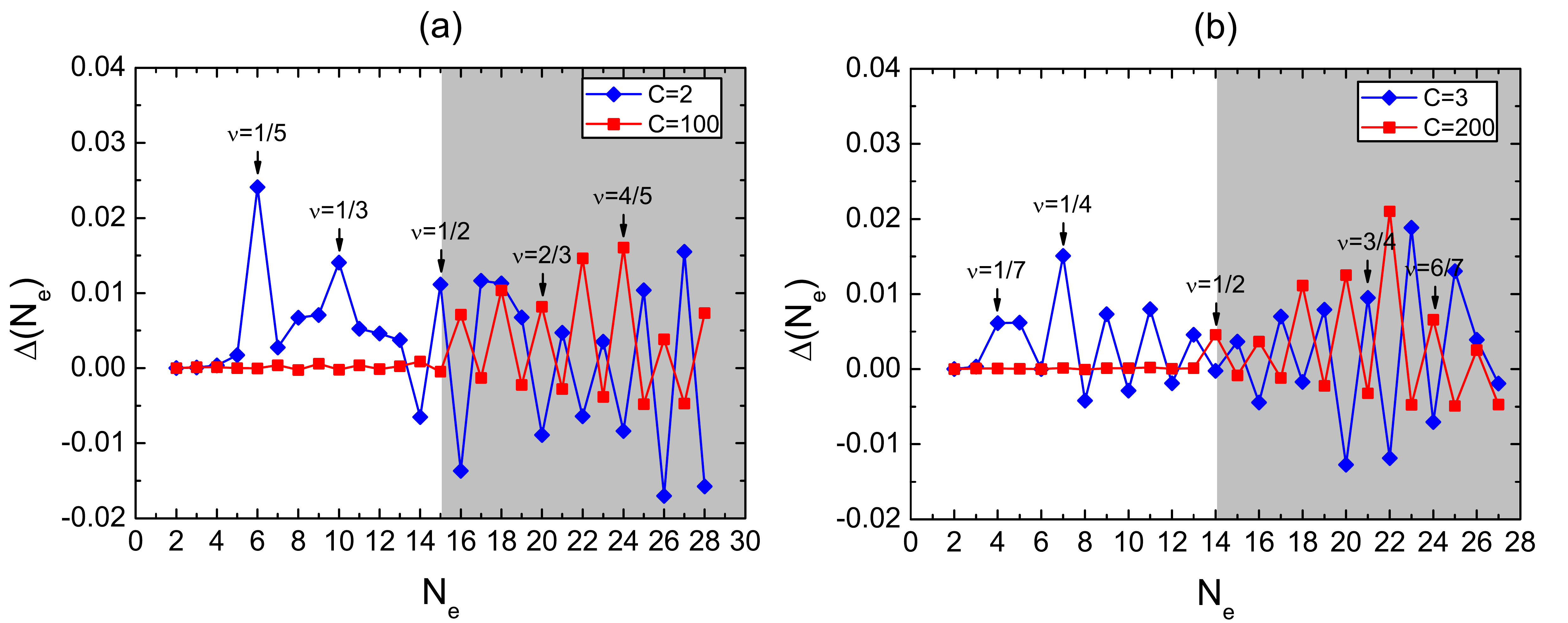}
\caption{The discontinuity of chemical potential, $\Delta(N_e)$, plotted against electron number $N_e$ in the (a) $6\times5$ and (b) $7\times4$ lattice. The parameters are $t_1=-1,\lambda_1=1.1,t_2=\lambda_2=0$.}
\label{fig:compress}
\end{figure*}

\begin{figure}
\includegraphics[width=\linewidth]{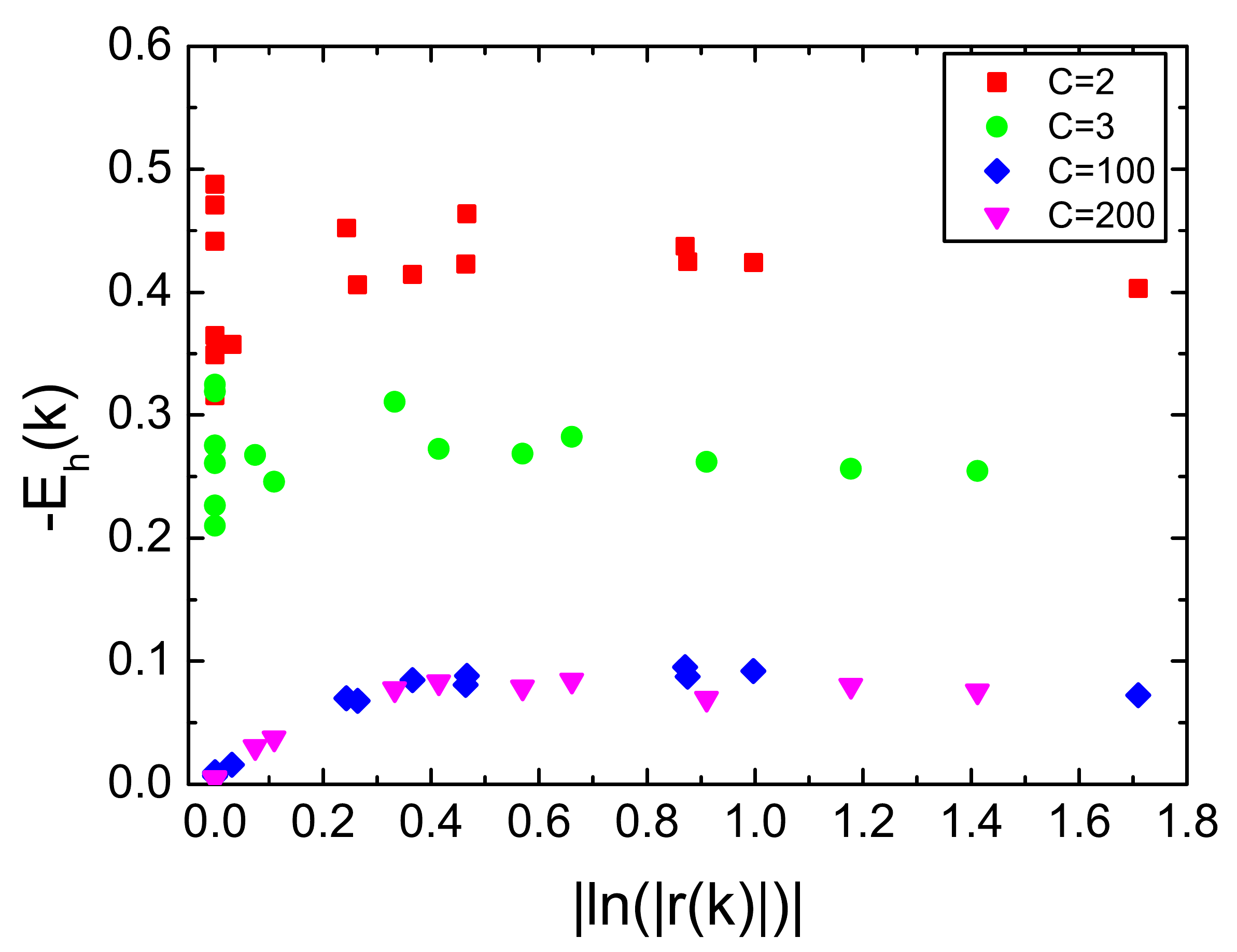}
\caption{$-E_h(\mathbf k)$ plotted against $|\xi(r(\bb k))|=|\log (|r(\bb k)|)|$ for ${\cal C}=2,100$ ($N_1\times N_2=6\times5$) and ${\cal C}=3,200$ ($N_1\times N_2=7\times4$), with parameters $t_1=-1,\lambda_1=1.1,t_2=\lambda_2=0$.
}
\label{fig:ehrk}
\end{figure}

To describe the main features of the phase diagram, we firstly note that the energy spectrum is symmetric under the transformation $\lambda\leftrightarrow-\lambda$ and $t_{\perp}\leftrightarrow-t_{\perp}$, so we can focus on the region $\lambda_1\geq0$ and $t_{\perp}\geq0$ without loss of generality.
Let us start with fixing $t_{\perp}=0$. In this case, the system is basically insulating at quarter filling, with a finite excitation gap.
However, there are two exceptional points: $\lambda_1=0$ and $\lambda_1=\sqrt{3}$, where the system becomes gapless.
These two points may be termed as quantum critical points, which separate two insulating regions.
The region of small $\lambda$: $0\leq\lambda_1\leq\sqrt{3}$ corresponds to the parameter space we focus on in the main text.
As increasing $t_{\perp}$, the system turns into a Weyl semi-metal (WS) in this region, signaled by the touching of lowest and second-lowest energy bands,
as is discussed in the main text.

More complicated changes can be observed for larger $\lambda_1$ region: $\lambda_1\geq\sqrt{3}$.
In this case, as increasing $t_{\perp}$, the energy gap closes indirectly. In other words, the top of lowest band and the bottom of second-lowest band exist at different momenta, and they exhibit a level crossing. This leads to the formation of Fermi surfaces with the same density of electron-like and hole-like carriers; the system enters a compensated metallic (CM) phase.

Within the CM phase, the direct gap also closes on increasing $t_{\perp}$, and the Weyl nodes show up,
even though the existence of bulk carriers masks the influence of small carrier density around Weyl nodes.
However, as increasing $t_{\perp}$ further, Fermi surfaces disappear, and another Weyl semi-metal phase appears.
This WS phase has a distinct feature that the Weyl cones are extremely tilted so that one of the conical generatrix of the Weyl cone becomes horizontal.
This leads to the line-like Fermi node, rather than Fermi points, and gives finite density of states at Fermi level, in contrast to the ordinary WS phase.
We term this WS phase as ``flat Weyl semi-metal", and distinguish it from the WS phase for small $\lambda_1$.

\subsection*{Gapless three-dimensional limit}
Below we will elaborate on the inability of interactions to open up a gap in the three-dimensional limit, ${\cal C}=N\rightarrow \infty$.

{\it Compressibility.---}
A useful quantity to consider is
\begin{equation}
\Delta(N_e)=N_e\Big[\frac{E_0(N_e+1)}{N_e+1}+\frac{E_0(N_e-1)}{N_e-1}-2\frac{E_0(N_e)}{N_e}\Big] \nonumber ,
\end{equation}
which (usually called charge gap) is a measure of (inverse) compressibility \cite{cooper}. Here $E_0(N_e)$ is the ground energy of a system with $N_e$ electrons. A clear hint of an incompressible state, such as a fractional Chern insulator, is the existence of a peak in $\Delta(N_e)$ at the pertinent filling fraction (assuming a fixed lattice size). In Fig. \ref{fig:compress} we plot  $\Delta(N_e)$ for a system with $N_1\times N_2=6\times 5$ unit cells and Chern number $\mathcal{C}=2,100$ (left panel) and  $N_1\times N_2=7\times 4$ and $\mathcal{C}=3,200$ (right panel). The lattice sizes are deliberately chosen as to accommodate filling fractions of known FCIs ($\nu=1/5$ for $\mathcal{C}=2$ and $\nu=1/7$ for the $\mathcal{C}=3$ case), but the general arguments given below for the difference between large and small $\mathcal{C}$ are independent of the particular lattice chosen. Perhaps somewhat surprisingly, our plots show a large number of peaks, also at high Chern numbers. Below, we corroborate the arguments given in the main text, and argue that, while the small $\mathcal{C}$ peaks frequently corresponds to FCIs, the large $\mathcal{C}$ peaks, which are instable against flux insertion, are due to finite size effects rather than corresponding to FCIs. In passing, we note that some of the small $\mathcal{C}$ peaks present may correspond to FCIs not reported previously, but a detailed investigation thereof is left for future studies.

{\it Hole-dispersion and penetration depth.---}
A striking feature of the compressibility plots (Fig. \ref{fig:compress}) is the particle-hole asymmetry. This is a direct consequence of the lack of translation invariance in the band. Following Ref.~\cite{andreasS}, this is readily seen by performing a particle-hole transformation, $c_{\mathbf k} \rightarrow c_{\mathbf k}^\dagger$ within the band. Focusing on fermions, the projected Hamiltonian transforms to
\begin{eqnarray}H&=& \sum_{\mathbf k_1,
 \mathbf k_2, \mathbf k_3, \mathbf k_4}V_{\mathbf k_1 \mathbf k_2 \mathbf k_3 \mathbf k_4}^{{\cal C}=N}c^\dagger_{\mathbf k_1}c^{\dagger}_{\mathbf k_2}c_{\mathbf k_3}c_{\mathbf k_4}\rightarrow
 \nonumber\\ &\rightarrow&\sum_{\mathbf k_1,
 \mathbf k_2, \mathbf k_3, \mathbf k_4}\Bigr(V_{\mathbf k_1 \mathbf k_2 \mathbf k_3 \mathbf k_4}^{{\cal C}=N}\Bigr)^*c^\dagger_{\mathbf k_1}c_{\mathbf k_2}^{\dagger}c_{\mathbf k_3}c_{\mathbf k_4}+\nonumber\\ &+&\sum_{\mathbf k}E_h(\mathbf k)
c^{\dagger}_{\mathbf k}c_{\mathbf k},\nonumber
\label{ph}
\end{eqnarray}
which generates an effective single-hole dispersion
\begin{eqnarray}E_h(\mathbf k)=\sum_{\mathbf m}(
 V^{{\cal C}=N}_{\mathbf m \mathbf k \mathbf m \mathbf k}+ V^{{\cal C}=N}_{\mathbf k \mathbf m \mathbf k \mathbf m}- V^{{\cal C}=N}_{\mathbf k \mathbf m \mathbf m \mathbf k}- V^{{\cal C}=N}_{\mathbf m \mathbf k \mathbf k \mathbf m}).\nonumber
 \end{eqnarray}
Within a Landau level, $E_h(\mathbf k)$ is dispersionless (i.e. a chemical potential) while $E_h(\mathbf k)$ is generically dispersive in a Chern band due to the breaking of translation invariance in reciprocal space.

That peaks in $\Delta(N_e)$ are only seen above $\nu=1/2$ for large ${\cal C}$ is a clear sign that $E_h(\mathbf k)$ is increasingly relevant for larger $\nu$. As argued in Ref.~\cite{andreasS} for the case of large filling fractions in flat ${\cal C}=1$ bands, a significant dispersion $E_h(\mathbf k)$ leads to gapless states; the ground state is simply obtained by filling the states with largest $E_h(\mathbf k)$, and such a state has Fermi surface(s) and supports gapless excitations in the thermodynamic limit. However, in a finite size lattice, this may result in sizable gaps which explain the peaks in $\Delta(N_e)$ occurring at large  ${\cal C}$ and high filling fractions.

This connects nicely to the argument given in the main text for the generic existence of gapless excitations for large ${\cal C}$ by the fact that $E_h(\mathbf k)$ tends to zero precisely when the penetration depth diverges (which occurs on the lines connecting $\Gamma$ and $M$). In Fig. \ref{fig:ehrk} we show precisely how $E_h(\mathbf k)$ correlates with the inverse correlation length  $|\xi(r(\bb k))|=|\log (|r(\bb k)|)|$ for Chern numbers ${\cal C}=N=2,3,100$ and $200$.

\begin{figure}
\includegraphics[width=\linewidth]{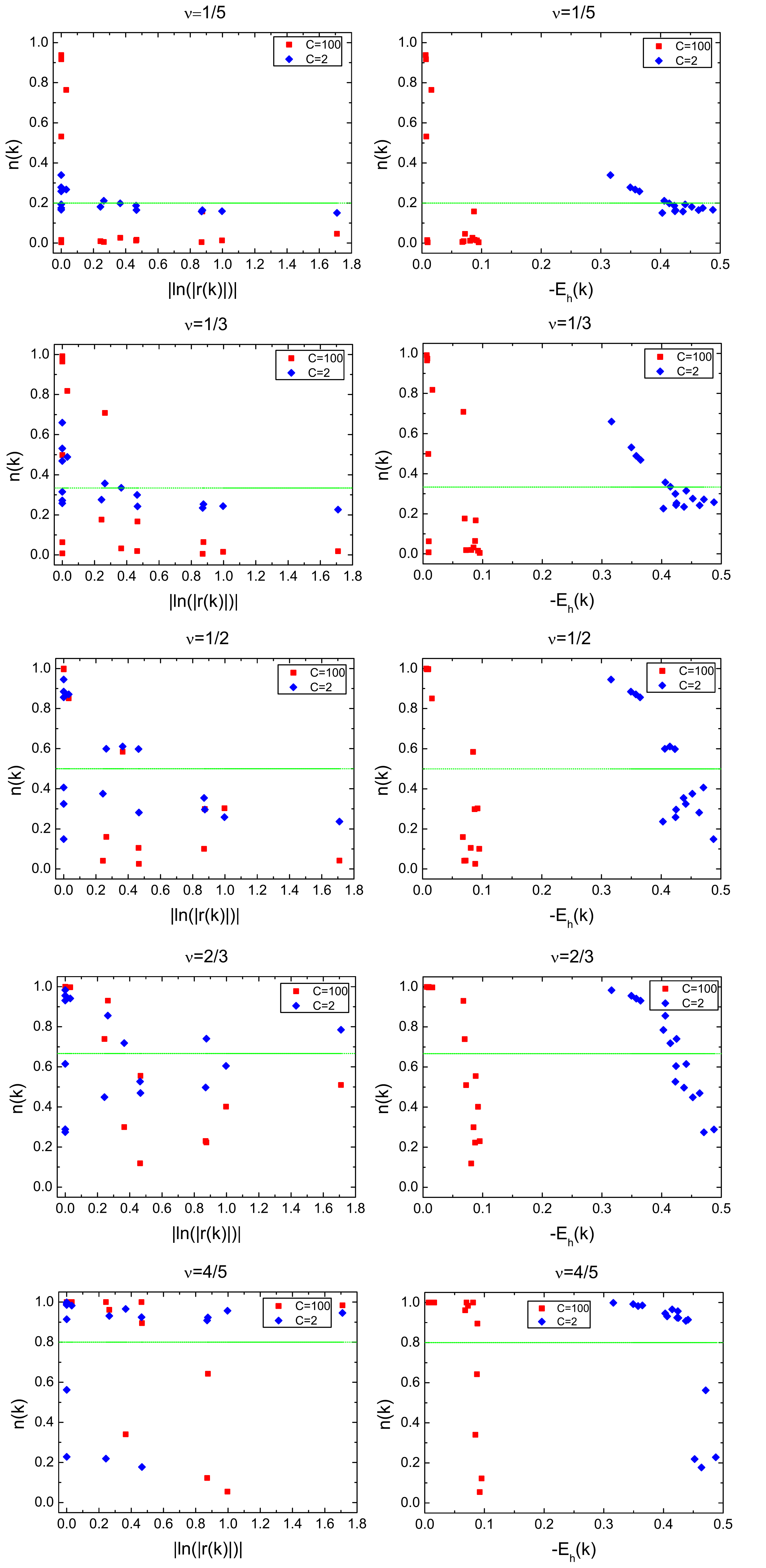}
\caption{$n(\bb k)$ plotted against $|\xi(r(\bb k))|=|\log (|r(\bb k)|)|$ (left column) and $-E_h(\bb k)$ (right column) for Chern numbers $C=2$ (blue diamonds) and $C=100$ (red squares) at various $\nu$ in the $6\times5$ lattice, with parameters $t_1=-1,\lambda_1=1.1,t_2=\lambda_2=0$. The green dotted lines indicate the reference $n(\bb k)=\nu$.}
\label{fig:nk2}
\end{figure}

\begin{figure}
\includegraphics[width=\linewidth]{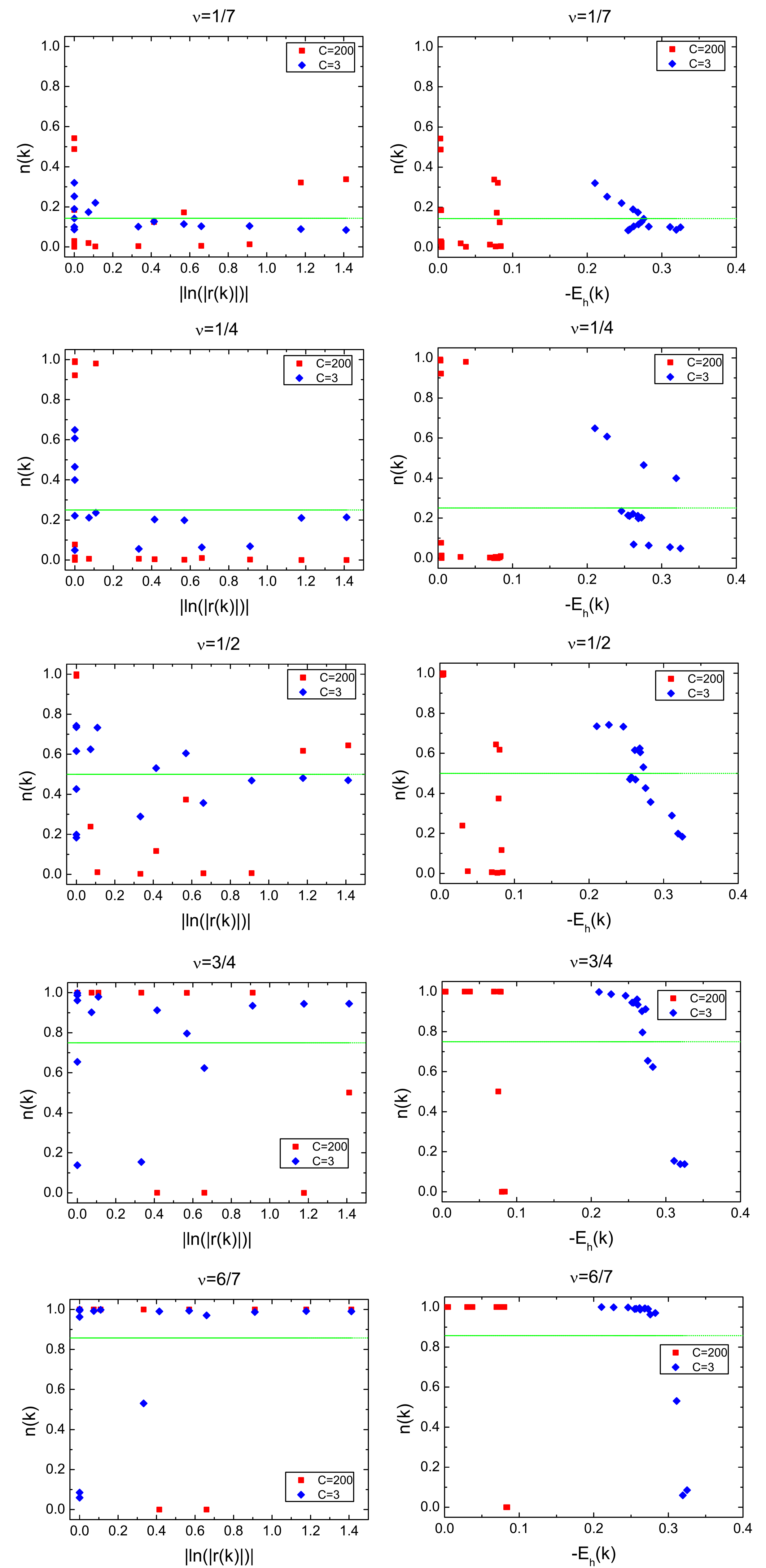}
\caption{$n(\bb k)$ plotted against $|\xi(r(\bb k))|=|\log (|r(\bb k)|)|$ (left column) and $-E_h(\bb k)$ (right column) for Chern numbers $C=3$ (blue diamonds) and $C=200$ (red squares) at various $\nu$ in the $7\times4$ lattice, with parameters $t_1=-1,\lambda_1=1.1,t_2=\lambda_2=0$. The green dotted lines indicate the reference $n(\bb k)=\nu$.}
\label{fig:nk3}
\end{figure}

{\it Occupation number distributions.---}
The arguments given above and in the main text can be further tested by numerically diagonalizing the full interacting many-particle problem in a small finite size sample and then plotting $n(\mathbf k)$ versus $|\xi(r(\bb k))|=|\log (|r(\bb k)|)|$ as well as $E_h(\mathbf k)$. In Figs. \ref{fig:nk2}  and \ref{fig:nk3}, we provide this for the system sizes, Chern numbers and filling fractions where we observe peaks in $\Delta(N_e)$ (cf. Fig. \ref{fig:compress}). We observe that $n(\mathbf k)$ is only smooth and approximately constant at small Chern numbers ${\cal C}$ and filling fractions $\nu$. At large $\nu$ we always find that $n(\mathbf k)$ for a Fermi surface like feature when plotted against $-E_h(\mathbf k)$; this holds true both for small and large Chern numbers. For small $\nu$ and large ${\cal C}$, we find that $n(\mathbf k)$ also correlates strongly with  $|\xi(r(\bb k))|$---it simply costs very little interaction energy to put further particles wherever $\xi(r(\bb k))$ diverges. That $n(\mathbf k)$ is never close to being homogenous at any filling fraction for ${\cal C}=100$ and $200$---in each case there are orbitals with $n(\mathbf k)\approx 0$ and/or $n(\mathbf k)\approx 1$---is strongly corroborating our conclusion that the flat bands at large ${\cal C}$ generically remain gapless as interactions are turned on.

\subsection*{Further evidence for the generalized ${\cal C}=2$ Moore-Read FCI}

\begin{figure*}
\includegraphics[width=\linewidth]{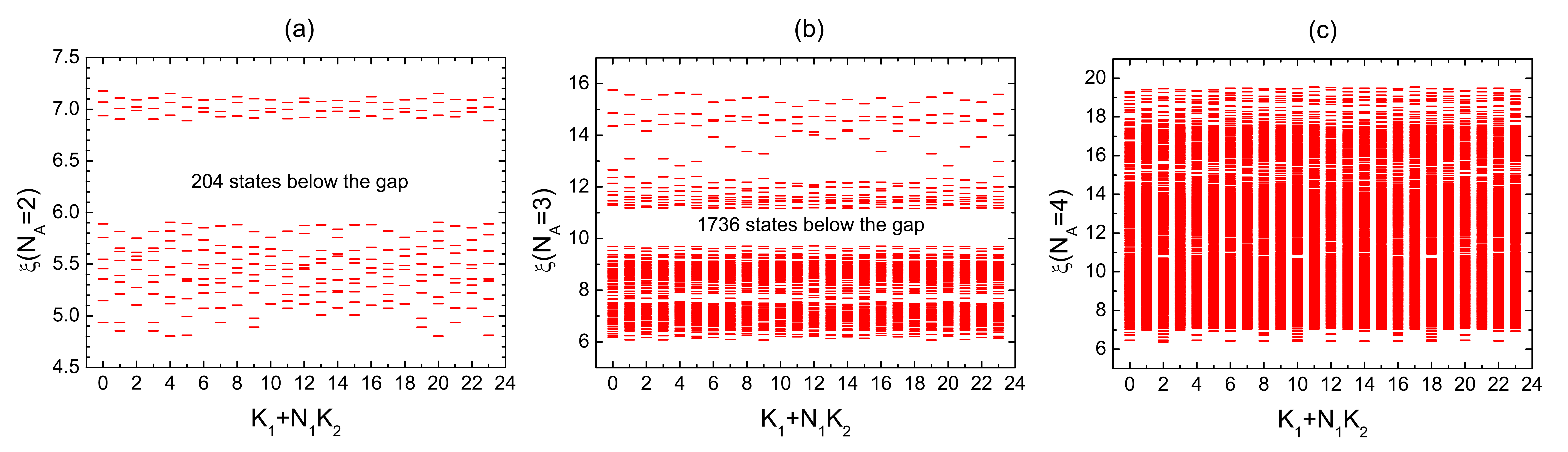}
\caption{The particle-cut entanglement spectrum for $N_e=8,N_1\times N_2=4\time6$ with three-body interaction at $\nu=1/3$ in ${\cal C}=2$ band.
The parameters are $t_1=-1,\lambda_1=0.9,t_2=\lambda_2=0$.}
\label{fig:PES}
\end{figure*}

{\it Entanglement spectrum.---}
In order to further elucidate the FCI character of $\nu=1/3$ states stabilized by three-body interaction in ${\cal C}=2$ band, we calculate the particle-cut
entanglement spectrum (PES) \cite{rbprx}. By dividing the system into $N_A$ particles and $N_e-N_A$ particles, the information of the quasihole excitation is encoded in the spectrum of the reduced density matrix of part $A$. In Fig.~\ref{fig:PES}, we show the PES for $N_e=8$ electron with $N_A$ from $2$ to $N_e/2$. The fact that a number of levels exist below a clear gap in the PES for small $N_A$ is sufficient to rule out the possible competing non-topological states such as charge density waves. Although the gap is invisible for the largest $N_A$, this is a situation often encountered in finite size studies, even in the case of very robust topological phases. Most notably, the PES gap is also absent even for the Coulomb ground state at $\nu=1/3$ in the lowest Landau level \cite{pes}, which is a well-known fractional quantum Hall state in the Laughlin phase.

The PES data together with the topological degeneracy and spectral flow in the main text strongly support that the ground states are FCIs. The excitations of this phase are most likely non-Abelian as indicated by the high ground state degeneracy, and most saliently, by the local three-body constraint they originate from which is closely analogous to that of ordinary ${\cal C}=1$ Moore-Read fractional quantum Hall states \cite{gww} and their lattice analogues \cite{nonab1,nonab2}. In this context we also note that a repulsive two-body interaction leads to another candidate FCI state at  $\nu=1/3, {\cal C}=2$, which has also not been reported earlier, albeit with a three-fold degeneracy and, most likely, Abelian low energy excitations.

\end{document}